\documentclass[english]{aastex}
\usepackage[T1]{fontenc}
\usepackage[latin9]{inputenc}
\usepackage{geometry}
\usepackage{amssymb}
\usepackage{graphicx}

\pdfoutput=1
\makeatletter

\usepackage{amssymb,amsmath}

\makeatother

\usepackage{babel}
\begin{document}

\title{Bulk Composition of GJ 1214b and other sub-Neptune exoplanets}

\author{Diana Valencia }

\affil{Earth, Atmosphere and Planetary Sciences Department, Massachusetts
Institute of Technology, 77 Massachusetts Ave, Cambridge, MA, 02139}

\email{dianav@mit.edu}

\author{Tristan Guillot }

\affil{Université de Nice-Sophie Antipolis, CNRS UMR 7293, Observatoire
de la Côte d'Azur, BP 4229, 06300 Nice Cedex 4, France}

\author{Vivien Parmentier }

\affil{Université de Nice-Sophie Antipolis, CNRS UMR 7293, Observatoire
de la Côte d'Azur, BP 4229, 06300 Nice Cedex 4, France}

\and{}

\author{Richard S. Freedman}

\affil{Seti Institute, Mountain View, CA \& Nasa Ames Research Center MS
245-3 P.O. Box 1 Moffett Field, CA 94035-0001}
\begin{abstract}
GJ1214b stands out among the detected low-mass exoplanets, because
it is, so far, the only one amenable to transmission spectroscopy.
Up to date there is no consensus about the composition of its envelope
although most studies suggest a high molecular weight atmosphere.
In particular, it is unclear if hydrogen and helium are present or
if the atmosphere is water dominated. Here, we present results on
the composition of the envelope obtained by using an internal structure
and evolutionary model to fit the mass and radius data. By examining
all possible mixtures of water and H/He, with the corresponding opacities,
we find that the bulk amount of H/He of GJ1214b is at most 7\% by
mass. In general, we find the radius of warm sub-Neptunes to be most
sensitive to the amount of H/He. We note that all (Kepler-11b,c,d,f,
Kepler-18b, Kepler-20b, 55Cnc-e, Kepler-36c and Kepler-68b) but two
(Kepler-11e and Kepler-30b) of the discovered low-mass planets so
far have less than 10\% H/He. In fact, Kepler-11e and Kepler-30b have
10-18\% and 5-15\% bulk H/He. Conversely, little can be determined
about the H$_{2}$O or rocky content of sub-Neptune planets. We find
that although a 100\% water composition fits the data for GJ1214b,
based on formation constraints the presence of heavier refractory
material on this planet is expected, and hence, so is a component
lighter than water required. A robust determination by transmission
spectroscopy of the composition of the upper atmosphere of GJ1214b
will help determine the extent of compositional segregation between
the atmosphere and envelope.
\end{abstract}


\section{Introduction}

The first step towards characterizing a planet is to infer its composition,
which can only be done if at least its mass and radius are known.
Within the realm of low-mass exoplanets, or super-Earths ($M<10\, M_{E}$),
there is now a handful of them with measured radii and masses. From
internal structure models, the interpretation of the data shows two
types of discovered planets: the rocky exoplanets including the high-density
ones CoRoT-7b and Kepler-10b, with a composition similar to that of
Mercury \citep{Valencia:CoRoT7b,Wagner:SE:12}, and Kepler-36b with
an Earth-like composition, and the ''volatile'' planets GJ 1214b,
the Kepler-11 system, 55 Cnc-e, Kepler-18b, Kepler-36c, Kepler-68b
and Kepler-30b that are too big to be made of rocks, as well as Kepler-20b
which sits at the boundary between necessarily volatile rich and possibly
rocky. These last assessments come from comparing their size to the
radius of planets made of the lightest rocky composition, one devoid
of iron (ie. a planet made of magnesium silicate oxides, MgO+SiO$_{2}$).
In addition, all these planets have effective temperatures that would
preclude an icy composition ($T_{eq}>300$ K). Thus, it is clear that
the volatile planets have gaseous envelopes. What remains to be determined
is the nature of this envelope. In particular, it is important to
assess if there is hydrogen and helium as this would mean that these
planets formed while the protoplanetary nebula was still around. 

Several studies have looked at the problem of inferring the bulk composition
of these planets, including their envelopes, through internal structure
models. However, the implementation of the opacities so far has been
too simple to carry out a consistent and systematic comparison between
volatile compositions. These studies have taken into account the effect
of composition in density (and entropy) via the equation of state
(EOS), but not in the values for the opacities. The main reason for
this shortcoming is that available opacity tables exist only at discrete
metallicity values. Because the evolution of gaseous planets towards
contraction depends on how opaque or transparent the atmosphere is,
deconvolving composition and opacity values may cause an over or under
estimation of the final radius of the planet, skewing the interpretation
of the data. In view of this problem, we focus on obtaining an analytical
fit to the discrete Rosseland opacity tables that would allow us to
interpolate to any composition spanning a hydrogen/helium + water/ices
composition for the envelope. 

In this study we focus our attention on GJ1214b, and compare its bulk
composition to the other volatile planets, because it is the first
low-mass planet with a measured spectrum and hence with an estimate
of the composition of the upper atmosphere. Due to its size relative
to its host star and the fact that the system is close enough to be
bright, this planet is amenable to transmission spectroscopy. So far,
several groups have obtained data at different wavelengths leading
to a rough spectrum of GJ 1214b. \citet{Bean:GJ1214b,Bean:GJ1214b:2011,Desert:GJ1214b:2011,Crossfield:GJ1214b:2011,Berta:GJ1214b:2011,Fraine:GJ1214b:13}
have all suggested a water-dominated atmosphere or hazes to explain
the featureless spectra they obtain, while \citet{Croll:GJ1214b:2011,deMooij:GJ1214b:2012}
suggested a low-molecular weight atmosphere. One caveat of these studies
is that the inferences depend on small differences between the data
and the one-sigma level uncertainty of the atmospheric compositional
models. Increasing the uncertainty two-fold would greatly impair the
inference of atmospheric composition.

On the other hand, internal structure models can help constrain the
bulk composition of a planet and thus complement the results from
transmission spectroscopy. Two previous studies have investigated
the composition of GJ 1214b. \citet{Rogers_Seager:GJ1214b} proposed
three different compositions and their respective origin for the envelope
of GJ 1214b: a primordial hydrogen and helium envelope acquired while
the protosolar nebula was still around, a water envelope acquired
in ice form with subsequent evaporation, or a hydrogen envelope which
was outgassed from the rocky interior. They used the opacity values
by \citet{Freedman:opacities:2008}, and a static model (no contraction
from the envelope) based on the parameterized grey atmospheric model
by \citet{Guillot:radiat:2010}. \citet{Nettelmann:GJ1214b:2011}
considered the composition of GJ 1214b to be a mixture of H/He and
water, with varying proportions of the two. They have an evolution
model that considers cooling and contraction of the envelope, a non-grey
atmospheric model and opacities that are 50 times solar. We add to
the discussion by using an internal structure model that improves
on the implementation of the opacities, and a comprehensive study
of the possibilities for the composition of the planet. 

In section 2 we describe the structure model used, and the implementation
of the opacities. In section 3 we show the results for GJ 1214b and
compare them to the other volatile transiting low-mass planets. Finally
we present our summary and conclusions in section 4.

\section{Model}

\subsection{Structure and Equation of State}

We treat planets as differentiated objects with an Earth-like nucleus
below an envelope composed of hydrogen and helium (H/He) as well as
water (H$_{2}$O). We use the combined internal structure model of
\citet{Valencia_et_al:2006} for the Earth-like nucleus (with composition
33\% by mass iron core + 67\% magnesium silicate mantle with 10\%
iron by mol - {[}(Mg$_{0.9}$, Fe$_{0.1}$)SiO$_{3}$ +(Mg$_{0.9}$,
Fe$_{0.1}$)O{]}) and CEPAM numerical model \citep{CEPAM:1995} for
the gaseous envelope. The two are tied at the solid surface by ensuring
continuity in mass and pressure. At this point we have not imposed
continuity in the temperature justified in part by the small effect
of temperature in the density of rocks. 

The EOS used is the Vinet EOS \citep{Vinet:1989} for the rocky interior
by combining the EOS of the end members with the additive density
rule to obtain an EOS of the mixture that is then used in the integration
of the structure equations. For the envelope we use the EOS of \citet{Saumon:EOS:1995}
for hydrogen and helium, considering always a fixed proportion of
Y=0.27 by mass of helium to the total amount of H$_{2}$ + He. For
the water, we combine the EOS of \citet{French:water:2010} that is
relevant for temperatures above 1000 K with the NIST EOS \citep{NIST}
which is well suited for low temperatures, to span the temperature
range between the critical point of water and 10,000 K.

\subsection{Opacities}

Owing to the fact that we are interested in constraining the composition
of the envelope by spanning all possible combinations between the
end members H/He and H$_{2}$O, we need corresponding opacities. Unfortunately,
the data available for opacities is limited to a few discrete compositions.
It is also limited in its maximal pressure, implying that interior
models must rely (often implicitly) on extrapolations. We use the
data from \citet{Freedman:opacities:2008} updated to include revised
collisional induced absorption by H$_{2}$ molecules for a solar composition,
2 and 1/2 times solar composition, plus an additional two data sets
at 30 and 50 times solar (hereafter F08). We obtain an analytical
fit to the Rosseland opacities by using a non-linear least squares
minimization approach useful within the temperature and pressure ranges
relevant for planetary interiors. The data sets span temperatures
between 75 and 4,000 K, and pressures between 10$^{-6}$ to $300$
bars, and the fit extrapolates smoothly in pressure, temperature (see
Fig 1) and metallicity.

\begin{figure}
\begin{centering}
\includegraphics[width=0.9\textwidth]{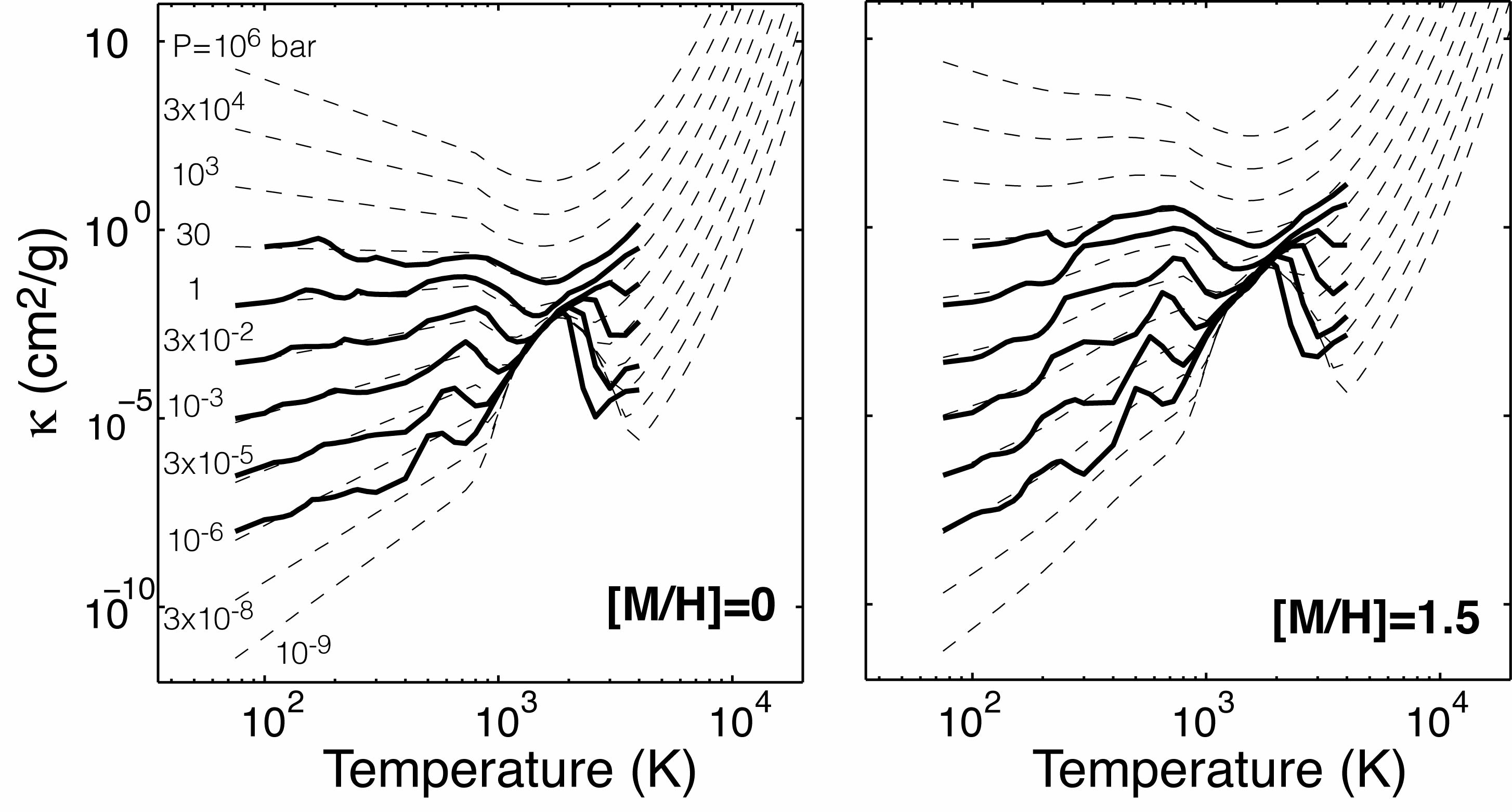}
\par\end{centering}

\caption{Opacity fit. The comparison between the Rosseland opacity data by
\citet{Freedman:opacities:2008} (solid lines) and our analytical
fit (dashed lines) is shown for solar metallicity (left) and a metallicity
30 times higher than solar (right). The extrapolation to low and high
pressures, as well as large temperatures is smooth. \label{fig:kappa_fit}}
\end{figure}

The analytical fit has the form for the opacity $\kappa_{\mathrm{gas}}$
(in g cm$^{-2}$) 

\begin{align}
\kappa_{\mathrm{gas}} & =\kappa_{\mathrm{lowP}}+\kappa_{\mathrm{highP}}\\
\log_{10}\kappa_{\mathrm{lowP}} & =c_{1}\left(\log_{10}T-c_{2}\log_{10}P-c_{3}\right)^{2}+\left(c_{4}\mathrm{\, met}+c_{5}\right)\\ 
\begin{split}
\log_{10}\kappa_{\mathrm{highP}} & =\left(c_{6}+c_{7}\log_{10}T+c_{8}\log_{10}T^{2}\right)+\log_{10}P\,\left(c_{9}+c_{10}\log_{10}T\right) \\
& \quad +  \mathrm{\,met}\, c_{11}\,\left(\frac{1}{2}+\frac{1}{\pi}\arctan\left(\frac{\log_{10}T-2.5}{0.2}\right)\right)
\end{split}
\end{align}

where $T$ is temperature in kelvins, $P$ is pressure in dyn cm$^{-2}$,
and $\mathrm{met}$ is the metallicity with respect to solar in logarithmic
scale (i.e. $\mathrm{met}=[\mathrm{M}/\mathrm{H}]$). This fit effectively
transitions smoothly between two different functions that are relevant
at low ($\kappa_{\mathrm{lowP}}$) and high pressures ($\kappa_{\mathrm{highP}}$)
respectively. The values for the coefficients are shown in table 1.
Figure \ref{fig:kappa_fit} shows a comparison between the data from
F08 for a solar composition and a metallicity 30 times higher ({[}M/H{]}=1.5)
and the results from our proposed analytical fit. 

\begin{deluxetable}{cccccc}
\tablewidth{0pt}
\tablecaption{Coefficients for Opacity Fit}
 \tablenum{1}
 \tablehead{\colhead{} & \colhead{all T} & \colhead{} & \colhead{} & \colhead{ T < 800 K} & \colhead{ T > 800 K }} 

 \startdata 
$c_1$ & -37.50  & & $c_6$  & -14.051 & 82.241  \\ 
$c_2$ & 0.00105 & & $c_7$  & 3.055   & -55.456 \\ 
$c_3$ & 3.2610  & & $c_8$  & 0.024   & 8.754   \\ 
$c_4$ & 0.84315 & & $c_9$  & 1.877   & 0.7048  \\ 
$c_5$ & -2.339  & & $c_{10}$ & -0.445  & -0.0414 \\ 
      &         & & $c_{11}$ & 0.8321  & 0.8321  \\  

\enddata


\end{deluxetable} 

The calculations by F08 consider a grain-free atmosphere with a composition
that evolves depending on the condensates that form and get removed
from the gaseous phase. We are interested in assessing whether or
not grains have an impact on the inference in envelope composition.
We focus on the end-member case of grains not settling into clouds,
but remaining mixed within the background gas. To model this type
of grains we turn to the calculations by \citet{Alex&Ferguson:opa:1994}
(hereafter AF94) to come up with a simple prescription that includes
the opacity contribution from mixed grains. \citet{Alex&Ferguson:opa:1994}
examined opacities at warm to high temperatures (between 700-12,500
K), and low density values captured in $\log_{10}\bar{R}=-7$ to +1
where $\bar{R}=\rho/T_{6}^{3}$, $\rho$ is the density in g cm$^{-3}$
and $T_{6}$ is the temperature expressed in millions of degrees (corresponding
densities are $10^{-15}$ to $10^{-8}$ g cm$^{-3}$ at 1,000 K and
$10^{-13}-10^{-5}$ g cm$^{-3}$ at 10,000 K) most relevant to the
conditions of the protoplanetary nebula. In contrast, planetary interiors
have larger density/pressure so that typical values are $+6<\log_{10}\bar{R}<+7$
. Their results show that grains are only present below a certain
temperature, which depends on the value of $\bar{R}$. Despite the
fact that AF94's data is calculated at very low values of $\bar{R}$,
there is a clear trend on the effect of grains that we extrapolate
to larger values of $\bar{R}$ (see Fig. 6 in \citet{Alex&Ferguson:opa:1994}).
We fit a simple linear trend within the regions where grains are present
and add this to the gas opacity:

\begin{equation}
\kappa = \begin{cases}
\kappa_{\mathrm{gas}}+\kappa_{\mathrm{grains}} & \text{ if $T<T^*_1$}, \text{ and} \\
\quad \quad \log_{10}\kappa_{\mathrm{grains}}=0.430 +1.3143\left(\log_{10}T-2.85\right) \\
\kappa_{\mathrm{gas}} & \text{ if $T>T^*_2$}
\end{cases}
\end{equation}

where $\log_{10}T_{1}^{*}=0.0245\log\bar{R}+1.971$ and $\log_{10}T_{2}^{*}=0.0245\log\bar{R}+3.221$.
The region between these two critical points is just a linear interpolation
between $\kappa_{\mathrm{gas}}(T_{1}^{*})+\kappa_{\mathrm{grains}}(T_{1}^{*})$
and $\kappa_{\mathrm{gas}}(T_{2}^{*})$. 

\begin{figure}
\begin{centering}
\includegraphics[width=1\textwidth]{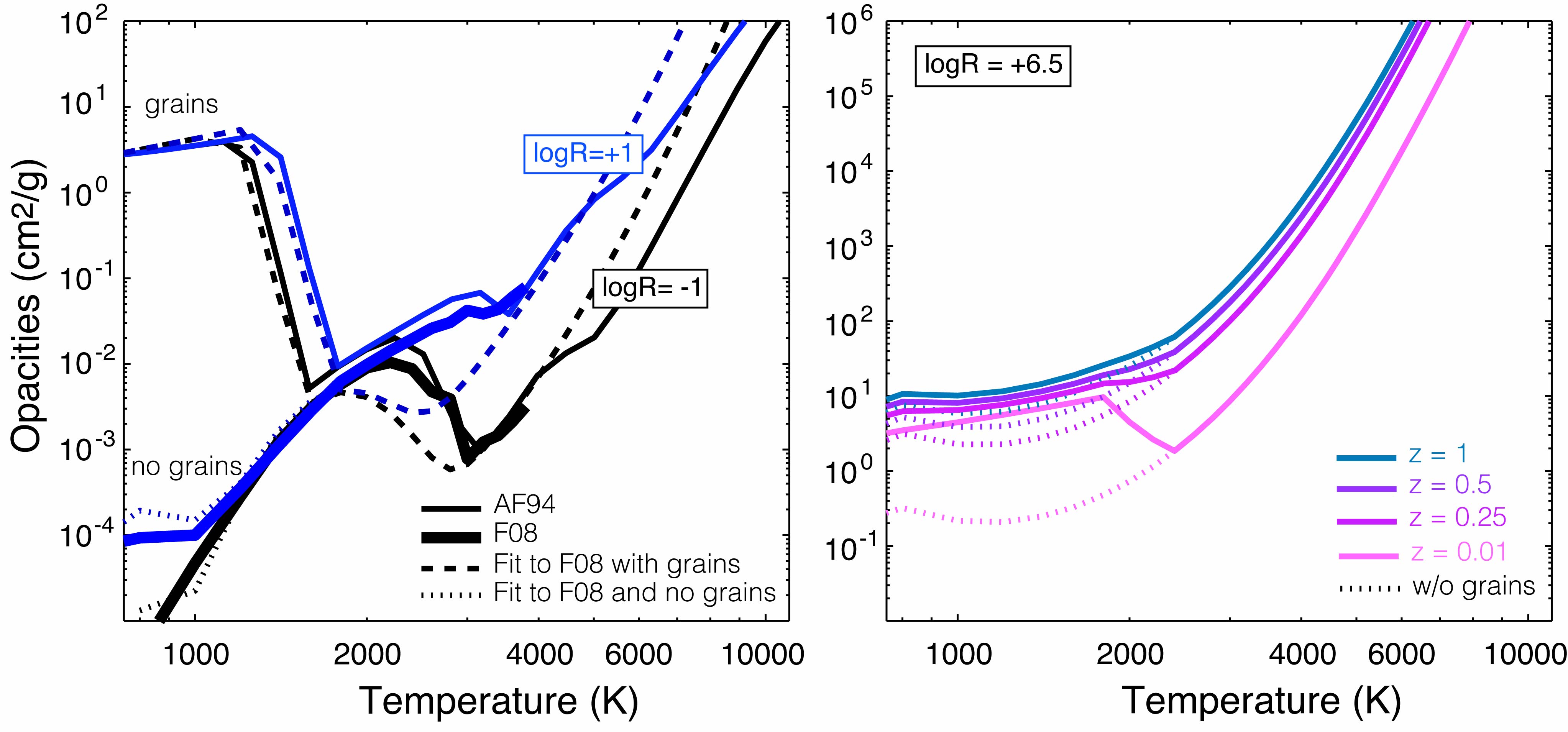}
\par\end{centering}

\caption{Comparison Opacities. Left) For a solar composition opacities according
to \citet{Alex&Ferguson:opa:1994} are shown in thin solid lines,
according to \citet{Freedman:opacities:2008} in thick solid lines,
our fit with grains in dashed lines and without grains in dotted lines
at $\log_{10}\bar{R}=-1$ in black and $\log_{10}\bar{R}=+1$ and
blue. Right) Comparison for three different envelope compositions:
solar, Z$_{\mathrm{ices}}$=0.01, (pink); 50\% ices + 50\% H/He ,
Z$_{\mathrm{ices}}$=0.5 (purple); and pure ices, Z$_{\mathrm{ices}}$=1,
(blue) for $\log_{10}\bar{R}=+6.5$ (relevant to planetary interiors)
and grainy (solid) and grain-free (dotted) cases. Our fit is an extrapolation
of the data beyond 4000 K, 300 bars and Z$_{\mathrm{ices}}$=0.25.}
\label{fig:opacities}
\end{figure}

We show the comparison between the two data sets (from AF94 and F08)
and our fit to the data with a prescription for grains at low temperatures
(dashed lines) and without grains (dotted lines) on Fig. \ref{fig:opacities}.
On the left, we compare the data (thin lines for AF94, thick lines
for F08) and our fit (dashed) for low densities ($\log_{10}\bar{R}=-1$
(black) and $\log_{10}\bar{R}=+1$ (blue)) and a solar composition.
According to AF94, the majority of the opacities for temperatures
lower than $T_{2}^{*}$ (or $\sim$1,800 K for $\log_{10}\bar{R}=+1$
) are due to grains, which we account for. The second feature of AF94
is a modest increase (a 'bump') in the opacities due to the presence
of water vapor at temperatures right above $T_{2}^{*}$ (between 1,800
K and 3,000 K for $\log_{10}\bar{R}=-1$ ), which becomes less prominent
with increasing value of $\bar{R}$ \citet{Alex&Ferguson:opa:1994}.
We note that this feature is missing in our fit to the data by F08
yielding differences in the opacities of almost an order of magnitude
around the $10^{-2}$ cm$^{2}$/g level within this high temperature
and low pressure ( e.g. low $\bar{R}$) range. However, this mismatch
we think may be less of an issue at pressure-temperature values pertinent
to planetary interiors given that the trend is for this feature to
be less prominent with increasing $\bar{R}$ values, and that the
opacities relevant for planetary atmospheres are in the 1-10$^{6}$
magnitude range. 

We compare the effect of envelope composition by showing in Fig \ref{fig:opacities}
(right) the opacities for a solar ( Z$_{\mathrm{ices}}$=0.01, where
Z$_{\mathrm{ices}}$ is the ratio of water/ices to envelope mass,
pink), a 50\% H/He + 50\% H$_{2}$O/ices mixture (Z$_{\mathrm{ices}}$=0.5,
purple) and a 100\% H$_{2}$O/ices envelope (Z$_{\mathrm{ices}}$=1,
blue) at a constant, more relevant value of $\log_{10}\bar{R}=+6.5$.
It can be seen that the opacities increase smoothly and monotonically
without grains (solid lines). In the presence of grains there is a
considerable (almost step-like) increase in opacities for temperatures
below $\sim$2,000K that depends on how much water there is, from
more than one order of magnitude for solar composition to just a few
tens of dex for envelopes rich in water/ices. The small effect of
grains on water-rich atmospheres is due to the fact that the opacities
are already quite high for such compositions.

We find that very quickly the opacities become high as soon as the
envelopes have non-negligible amounts of water/ices so that the difference
between opacities for a 50x solar envelope (or Z$_{\mathrm{ices}}$=0.25)
and a pure water/ices is only of the order of $\sim$50 g cm$^{-2}$
over a range that covers several orders of magnitude (see Fig. \ref{fig:opacities}). 

\paragraph{Extrapolation }

The pressure-temperature regime of super-Earths is between millibars
to a few megabars, and between a few hundred to $\sim$10,000 degrees
kelvin. The opacity database from which the fit is derived covers
this regime partially, and extrapolation is needed beyond 300 bars
and 4,000 K (Fig. \ref{fig:kappa_fit}). In fact, no current database
covers the planetary regime fully. Therefore internal structure studies
of planets use some sort of extrapolation. We used the work by AF94
to serve as a guide for the extrapolation to high temperatures, and
note that the opacities become increasingly more uncertain as the
pressures and temperatures increase much beyond the database of AF08.
This may not be too much of a problem as the high PT regime corresponds
to the deep interior of the planet (which is fully convective), and
most of the cooling is controlled by the radiative upper part of the
envelope/atmosphere (P < kbar, T< 2,000K) where the opacities are
either not too far from or within the database range. However it is
important to keep in mind the limitations of the extrapolation.

In general, the construction of high temperature databases requires
that transitions that originate in energy levels above the ground
state be included in the calculations. If this is not done, then there
will be missing opacity that will increase in magnitude as the temperature
increases. In the case of the opacity tables provided for this study,
wherever possible opacity data using \textquotedbl{}hot\textquotedbl{}
line lists has been used. These lists include line transitions from
levels that are not populated at room temperature so that the opacity
is more accurately represented at higher temperatures but these lists
still may not include all the opacity at the highest temperatures.
It is difficult to include in a quantum mechanical model all the levels
that may contribute opacity at the highest temperatures. 

By considering the species individually and assessing how they contribute
to the total opacity, it is reasonable to assume that the opacity
will continue to increase with T but only up to the point where the
population of the upper states reaches a limit where the effects of
additional increases in T are slight (due to the exponential dependence
of the Boltzmann factor). Since for most polyatomics the first excited
electronic state is far above the ground state, it is only the vibrational/rotational
levels in the ground electronic state that need to be considered.
This is especially true in the case of a main source of opacity, water,
where the first bound, excited state is $\sim$7.5 ev above the ground
state. On the other hand, an important source of absorption at high
T is the presence of free electrons and the associated free-free and
bound-free opacities, which may have a more significant effect than
the filling of the band gaps, counteracting the saturation effect
previously discussed.

In addition, the effect of very high pressure may also have a leveling-off
component. For moderate pressures the line width will increase with
pressure but eventually a limit is reached when the density approaches
a value where the gas starts to behave more like a liquid and the
line width no longer increases linearly with pressure. Unfortunately,
the actual line width at these high pressures is not at all well-known
and the simple theories for line shape are no longer valid, making
extrapolation rather uncertain. 

To test how sensitive the extrapolation is to high temperatures we
use a synthetic opacity fit with a much lower dependence on temperature
(parameter $c_{8}=5$). We consider two planets with a mass of 0.020
$M_{\mathrm{Jup}}$, $T_{eq}=500$ K and a core mass fraction of 50\%:
one with a 100\% water/ices envelope and one with a mixture of 90\%
H/He and 10\% water/ices. We find that the radiative-convective boundary
moves shallower by 10\% and 20\% in pressure at 10 Gyr, respectively,
and that the opacity values increase by 12\% and 17\% respectively.
Not enough to change the PT structure or total radius significantly
(by 0.2\% and 5\% respectively at 2.5 Gyr and by less than 1 part
in 1000 for both cases at 10 Gyr). This confirms that the PT regime
for the opacities is most important up to the pressures and temperatures
that include the radiative-convective transition, which for these
warm sub-Neptunes is < 5 kbar and < 2000 K. 

In addition, the database used (AF08) only spans a limited range of
envelope compositions: solar, -1/2x solar, 1/2x solar, 30x solar and
50x solar in gaseous form (no solids). Thus, extrapolation is needed
to cover the whole space from solar to water-rich envelopes. From
this set we observe that the dominant dependence of the logarithm
of the opacity with metallicity is a linear dependence beyond some
estimated temperature ($\sim$ 3000K). This simple fit (see Eq. 3)
captures the intuitive behavior that opacity increases with the number
of molecules present, while fitting the database well. It is of course,
too simple of an extrapolation to capture the details. We await actual
data at larger metallicity values to compare to our fit, especially
as more sub-Neptune planets are found. 

We hope that in the future there will be no need for extrapolation,
and encourage the expansion of opacity databases to higher PT and
water rich compositions important for modeling the structure of low-mass
exoplanets, in the meantime our proposed fit may serve as a starting
point.

\subsection{Metallicity and composition}

To use the opacity fit, we relate the composition of the envelope
to metallicity. We consider the envelope to be composed of H$_{2}$-He
and 'ices', where the ices are composed of water, ammonia and methane
(H$_{2}$O + NH$_{3}$ + CH$_{4}$), in the same proportions as in
the solar nebula. We implicitly assume that there are no rock forming
minerals that could bind to oxygen, so that the amount of water in
the envelope is reflected in the amount of oxygen atoms ($N_{\mathrm{O}}$),
and that the other ices are fixed by the solar ratios of carbon and
nitrogen to oxygen.

This means that $Z_{\mathrm{ices}}$, the amount of 'ices' by mass
is 

\[
\mathrm{Z_{ices}}=\frac{N_{\mathrm{O}}\mu_{\mathrm{H_{2}O}}+N_{\mathrm{C}}\mu_{\mathrm{CH_{4}}}+N_{\mathrm{N}}\mu_{\mathrm{NH_{3}}}}{N_{\mathrm{H}}\mu_{\mathrm{H}}+N_{\mathrm{He}}\mu_{\mathrm{He}}+N_{\mathrm{O}}\mu_{\mathrm{O}}+N_{\mathrm{C}}\mu_{\mathrm{C}}+N_{\mathrm{N}}\mu_{\mathrm{N}}},
\]
where $N_{i}$ and $\mu_{i}$ are the number of atoms and the molecular
weight of species $i$, respectively. We take constant the proportion
of He to the total amount of mass in the non-metallic portion (H$_{2}$+He)
and equal to $c=0.27$ (i.e. using the conventional notation: Y/(X+Y)=0.27
). Therefore, the metallicity can be expressed as
\[
\left(\frac{N_{\mathrm{O}}}{N_{\mathrm{H}}}\right)=\left(\frac{1}{1-c}\right)\frac{\mu_{\mathrm{H}}\:\mathrm{Z_{ices}}}{a-b\,\mathrm{Z_{ices}}}
\]

\[
10^{\mathrm{met}}=\left(\frac{N_{\mathrm{O}}}{N_{\mathrm{H}}}\right)/\left(\frac{N_{\mathrm{O}}}{N_{\mathrm{H}}}\right)_{\mathrm{solar}},
\]
where $a=\mu_{\mathrm{H_{2}O}}+\left(N_{\mathrm{\mathrm{C}}}/N_{\mathrm{O}}\right)\mu_{\mathrm{CH_{4}}}+\left(N_{\mathrm{N}}/N_{\mathrm{O}}\right)\mu_{\mathrm{NH_{3}}}$,
and $b=\left(\mu_{\mathrm{O}}-2c\frac{\mu_{\mathrm{H}}}{1-c}\right)+\left(N_{\mathrm{\mathrm{C}}}/N_{\mathrm{O}}\right)\left(\mu_{C}-3c\frac{\mu_{\mathrm{H}}}{1-c}\right)+\left(N_{\mathrm{N}}/N_{\mathrm{O}}\right)\left(\mu_{\mathrm{N}}-4c\frac{\mathrm{\mu_{H}}}{1-c}\right)$.
We used the values of $N_{\mathrm{\mathrm{C}}}/N_{\mathrm{O}}=0.501$,
$N_{\mathrm{N}}/N_{\mathrm{O}}=0.138$ and $(N_{\mathrm{\mathrm{C}}}/N_{\mathrm{O}})_{\mathrm{solar}}=4.898\times10^{-4}$
from \citet{Lodders:2003}. This means that our opacity fit spans
values for the metallicity from solar to 457 times solar ($\mathrm{met}=2.66$),
corresponding to Z$\mathrm{_{ices}}$=1.

\subsection{Atmospheric Model}

\begin{figure}
\begin{centering}
\includegraphics{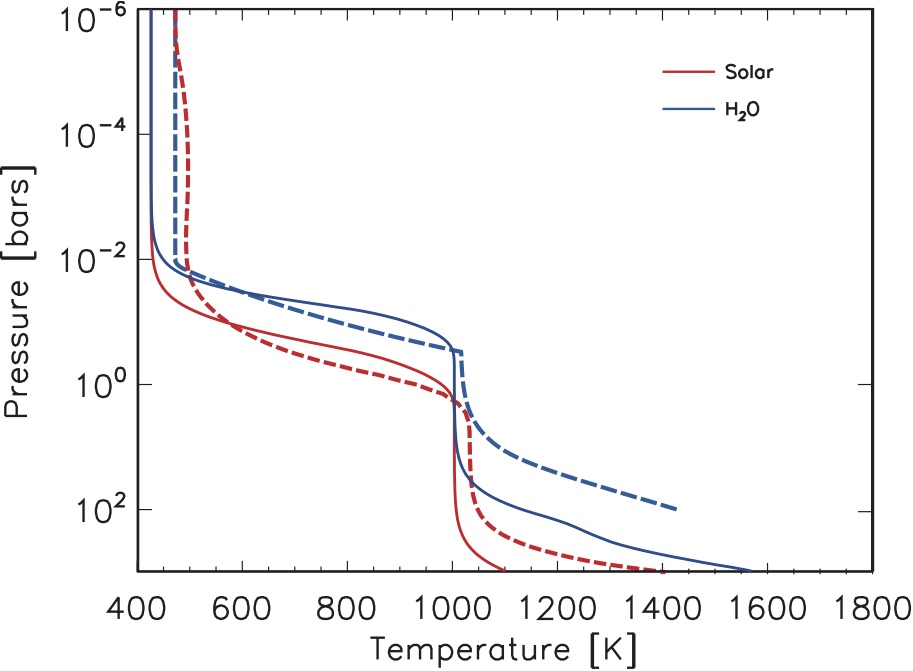}
\par\end{centering}

\caption{Upper atmosphere of GJ 1214b. Temperature pressure profiles from \citet{Miller-Ricci:GJ1214b}
(dashed lines) and from our model at 10 Gyr (solid lines) for a solar
composition atmosphere (red) and for a water atmosphere (blue).
\label{PTcomparison}}
\end{figure}

The upper boundary condition of our interior model is given by the
atmospheric model of \citet{Guillot:radiat:2010}. This analytical
model is valid for a plane-parallel atmosphere which transports both
a thermal intrinsic flux and a visible flux from the star. The visible
flux propagates downward from the top of the atmosphere and is absorbed
with an opacity $\kappa_{\mathrm{v}}$. The ratio of the visible to
the infrared opacities $\gamma\equiv\kappa_{\mathrm{v}}/\kappa_{\mathrm{IR}}$
is considered constant. Its value determines at which depth the radiative
energy from the star is deposited. For high values of $\gamma$, the
energy is deposited in the upper layers of the atmosphere and can
be lost toward space very easily. For values of $\gamma$ lower than
unity, the energy is deposited in deeper layers, where the atmosphere
is optically thick in the infrared. There, the energy cannot escape
the planet, and contributes to its global energy, slowing its contraction.
Theoretically, $\gamma$ could be calculated from the opacity tables.
However, we note that the temperature of the isothermal zone around
the 1 bar level is very sensitive to its value. Thus we use the value
of $\gamma$ that better reproduces the more sophisticated radiative
transfer models of \citet{Miller-Ricci:GJ1214b}. We choose $\gamma=0.032$,
which gives a temperature of 1,000 K around 1bar for GJ1214b, as can
be seen in figure \ref{PTcomparison}. \citet{Miller-Ricci:GJ1214b}
show that, in the case of GJ1214b, the temperature around 1 bar does
not depend strongly on the composition of the planet. Thus, we use
the same $\gamma$ for the different compositions considered. For
GJ 1214b, we find that the interior temperature is 62K at 0.1 Gyr,
40K at 1 Gyr, 35K at 2.5 Gyr and 24K at 10 Gyr for a solar atmosphere,
and 80K at 0.1 Gyr, 50K at 1Gyr, 42K at 2.5 Gyr and 35K at 10 Gyr
for a water-rich envelope.

Note that $\gamma$ will change with orbital distance. It is expected
to be higher for planets that are close-in. The Rosseland opacities
are calculated from the line by line opacities weighted by the Planck
function. Thus, planets with different equilibrium temperatures have
different values of the Rosseland thermal opacities. Changing the
equilibrium temperature by a small amount ($\sim O(100\,\mathrm{K})$)
will not change the position of the peak of the Planck function very
much. However, for hotter planets, not considered in this study, the
value of gamma could change significantly.

\section{Results}

\subsection{GJ 1214b}

We obtain the structure and total radius for planets with a mass of
5.09, 6.36, 7.63, 8.90 and 10.2 $\mathrm{M_{Earth}}$ (or 0.016, 0.020,
0.024, 0.028 and 0.032 $\mathrm{M_{Jup}}$) to span the mass of GJ1214b,
for different proportions of Earth-like nucleus to envelope, while
spanning all combinations of the end members H/He and H$_{2}$O for
the envelope. In other words we find a relationship between mass ($M$),
radius ($R$), Earth-like nucleus fraction to total mass (nf) and
proportion of water to total envelope mass (wf), in the form $R=R(M,\mathrm{nf},\mathrm{wf})$,
and spline interpolate in the three dimensions (mass, nf, wt). Since
we are interested in inferring the composition of a planet from its
transit radius, we consider the radius of the planet to be the height
at which the path traveled by the starlight would be equal to an optical
depth of unity. We examined three cases: (1) a grain-free envelope
and (2) a grainy envelope at an equilibrium temperature of $T_{eq}=500$
K and c) a grain-free envelope at $T_{eq}=600$ K. 

\begin{figure}
\begin{centering}
\includegraphics[width=1\textwidth]{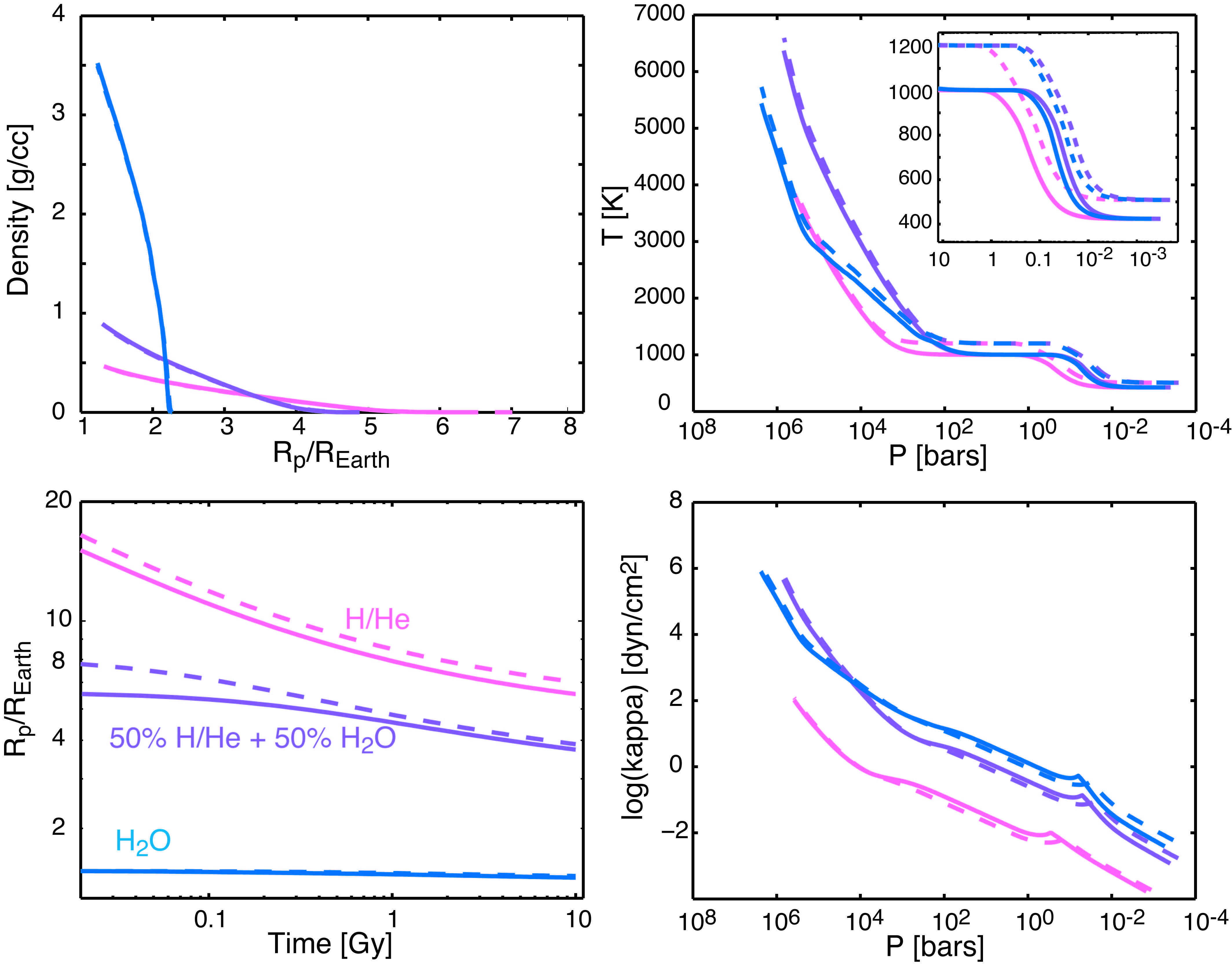}
\par\end{centering}

\caption{Envelope structure of a sub-Neptune. Top-left) Density structure,
top-right) pressure-temperature structure, bottom-left) evolution
(contraction) tracks, and bottom-right) opacity values for planets
that are made of 50\% by mass earth-like core below envelopes of different
compositions. Blue: 100\% H$_{2}$O envelopes; purple: 50\% H$_{2}$O
+ 50\% H/He; pink: 100\% H/He envelope. Solid and dashed lines correspond
to equilibrium temperatures of 500 and 600 K, respectively. The envelope
are grain-free in this case. The total mass of the planet is 0.020
M$_{\mathrm{Jup}}$.}
\label{fig:structure}
\end{figure}

Figure \ref{fig:structure} shows typical calculations for the planets
considered. In this case the planets have an earth-like nucleus that
makes up half of the planet's mass below envelopes of different compositions:
a) 100\% H$_{2}$O (blue), b) 50\% H$_{2}$O+ 50\% H/He (purple) and
c) 100\% H/He (pink). Starting from a high entropy state (corresponding
to $S=S(\chi_{\mathrm{env}},T_{10},P_{10})$, where $\chi_{\mathrm{env}}$
is the envelope's composition, $T_{10}$ and $P_{10}$ are the temperature
and pressure at 10 bars), the planets cool and contract according
to how much energy is being transported out (bottom left panel). The
solid and dashed lines correspond to equilibrium temperatures of 500
and 600 K. As it can be seen, this small difference  in equilibrium
temperature has little effect on the interior structure or evolution
of the planets. 

\begin{figure}
\begin{centering}
\includegraphics[width=0.9\textwidth]{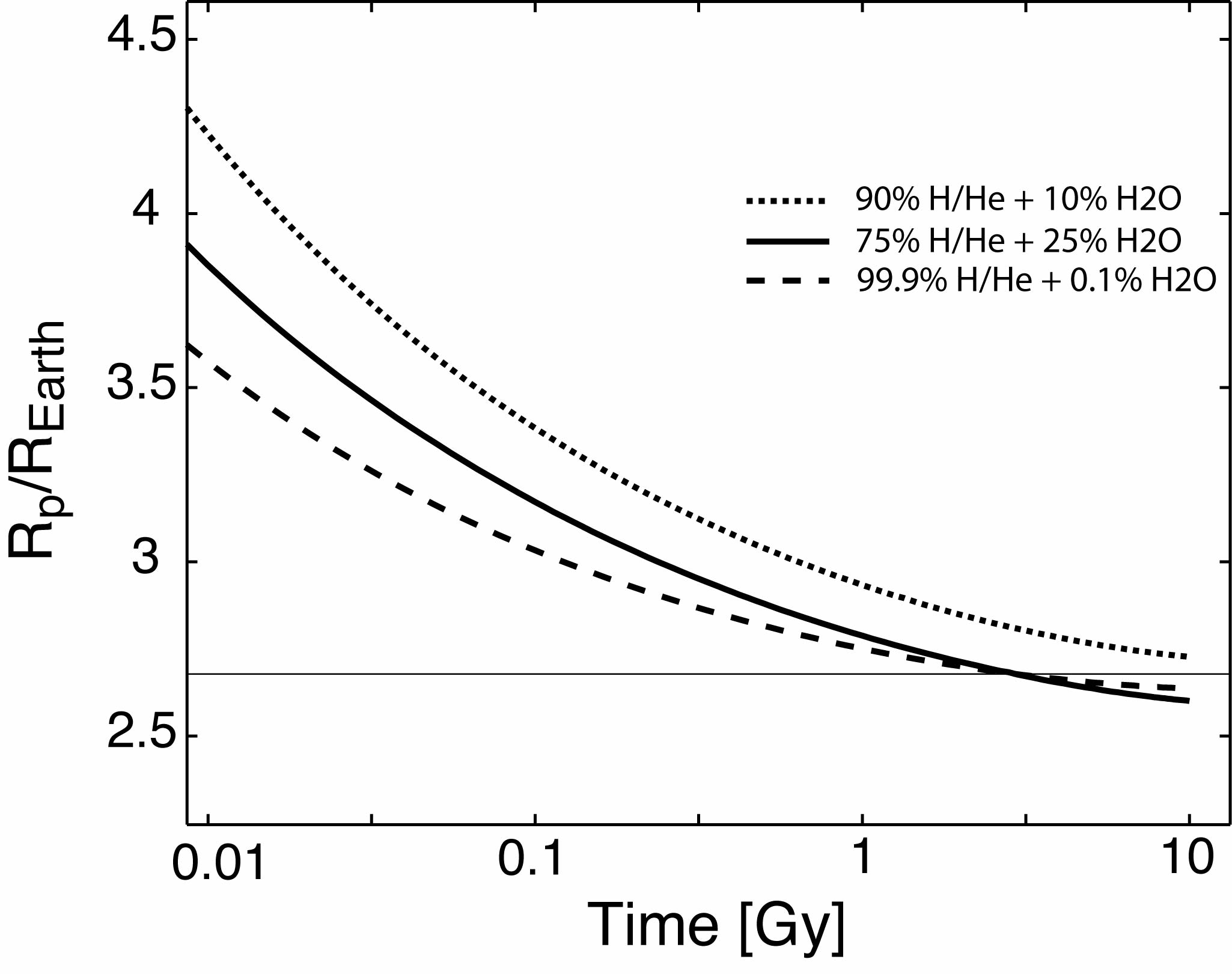}
\par\end{centering}

\caption{Degeneracy in envelope composition. Evolutionary tracks are shown
for a planet of mass 0.020 M$_{\mathrm{Jup}}$ and T$_{\mathrm{eq}}=$500
K, with an envelope that is 3\% by mass and different compositions:
75\% H/He + 25\% H$_{2}$O (solid line), 99.9\% H/He + 0.1\% H$_{2}$O
(dashed line) and 90\% H/He + 10\% H$_{2}$O (dotted lines). The latter
is shown for reference. The radius of 6.55 R$_{\mathrm{E}}$ is met
by the first two compositions at an age of $\sim$3 Gyr (fine horizontal
line).\label{fig:Degeneracy-in-envelope} }

\end{figure}

Not surprisingly, the envelopes that have lower molecular weight yield
the largest radii, while at the same time suffer the most contraction.
We find this trend to be true for most planets except for the ones
that have less than 10\% content of water in the envelope. That is
to say, that we find that planets that have envelopes of 100\% H/He
are slightly smaller than those that have 90\% H/He+10\% water/ices
envelopes. We attribute this to a competing effect between larger
envelope density that would make planets smaller for a given mass,
and higher opacities that slow down the cooling. At larger fractions
of water content in the atmosphere, the density effect dominates.
Interestingly, this effect gives rise to a new kind of degeneracy.
For the same value of envelope mass, two different combinations of
H/He + water/ices with two different evolutionary tracks, yield the
same radius at some given age (see Fig \ref{fig:Degeneracy-in-envelope}).
This illustrates the importance of using evolutionary models, as static
ones could miss these possibilities. By implementing the physics behind
contraction and evolution, the internal structure model is able to
resolve time-dependent possibilities. This degeneracy stands in contrast
to the one that arises from trade-offs between three or more compositional
end-members with different molecular weights -- iron cores, silicate
mantles, water/icy envelopes or oceans, H/He envelopes, which has
been readily identified \citep{Valencia_ternary,Adams_Seager_Tanton:2008,Rogers_Seager:2010a}.
The new degeneracy arises from differences in molecular weight and
thermo-physical properties (opacities) between water and H/He that
determine the cooling histories of the envelopes.

The age of GJ1214b is estimated to be between 3-10 Gyr \citep{GJ1214b},
which means the planet may contract considerably within this age range
adding another source of uncertainty when inferring the composition
of the envelope. The effect of contraction is most significant in
the early stages of evolution (< 1Gyr) and for H/He dominated envelopes,
and less important as the age of the planet increases or its envelope
is H$_{2}$O dominated. These two effects are shown in the bottom-left
of Fig. \ref{fig:structure}. To infer the composition of GJ1214b
we use a nominal age of 4.6 Gyr and then explore the effects of the
uncertainty in the age.

For the specific example shown in Fig \ref{fig:Degeneracy-in-envelope},
a planet with a mass of 0.020 $M_{\mathrm{Jup}}$, and an envelope
that makes up 3\% of the total mass, two different envelope compositions
yield the same radius of 6.55 R$_{\mathrm{E}}$ at 3 Gyr. An envelope
that is mostly H/He (99.9\% H/He and only 0.1\% water in the envelope)
that starts very expanded and contracts rapidly initially, and an
envelope that is made of 3/4 of H/He and 1/4 of water, that contracts
initially more slowly. To resolve this kind of degeneracy one would
need a radius measurement at two different ages, which is impossible
to obtain. Therefore, we find that for low-mass planets with a non-negligible
envelope or sub-Neptunes, there is an intrinsic and persistent degeneracy
that stems from the contraction history of the planet.

\begin{figure}
\begin{centering}
\includegraphics{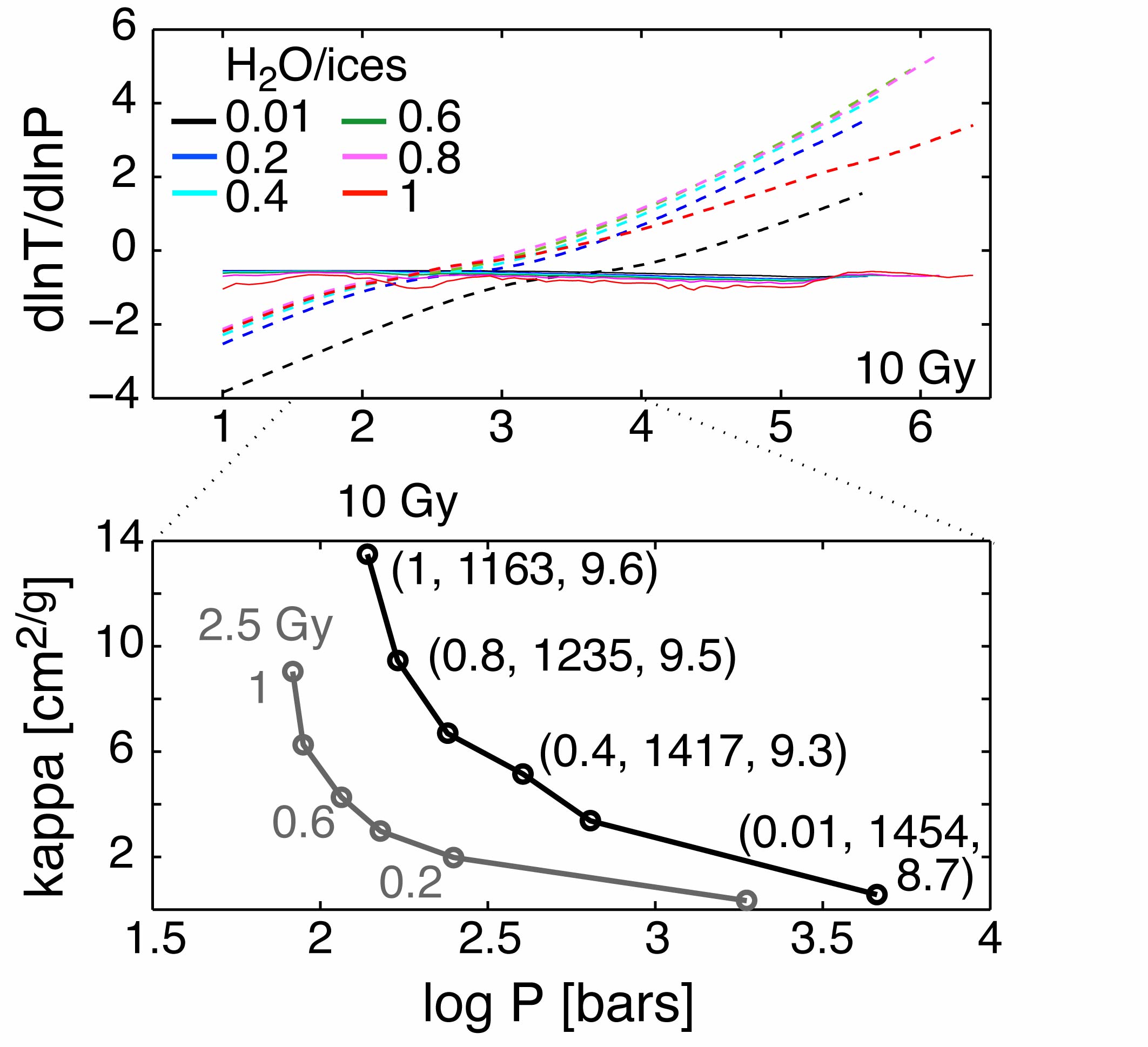}
\par\end{centering}

\caption{Adiabatic-Radiative boundary of a sub-Neptune with mass 0.020 M$_{\mathrm{Jup}}$
and T$_{\mathrm{eq}}=$500 K. Top) Adiabatic (solid lines) and radiative
(dashed) gradients at 10 Gyr of envelopes with compositions: solar
(black), 80\% H/He + 20\% H$_{2}$O/ices (blue), 60\% H/He + 40\%
H$_{2}$O/ices (cyan), 40\% H/He + 60\% H$_{2}$O/ices (green), 20\%
H/He + 80\% H$_{2}$O/ices (pink), 100\% H$_{2}$O/ices (red), over
an Earth-like nucleus that makes 50\% of the planet by mass. The region
where the radiative gradient is lower than the adiabatic one, the
planet loses heat via radiation. Bottom) The pressure (depth) and
corresponding opacity of the radiative-convective boundary for planets
of 2.5 Gyr (grey) and at 10 Gyr (black). The labels correspond: to
the proportion of H$_{2}$O/ices in the envelope, and the temperature
and entropy (in log) of the radiative-convective boundary. For 2.5
Gyr these values are (0.01, 1478, 8.7), (0.2, 1438, 9.1), (0.4, 1391,
9.3), (0.6, 1357, 9.4), (0.8, 1278, 9.5), (1, 1211, 9.6); and for
10 Gyr: (0.01, 1454, 8.7), (0.2, 1452, 9.1), (0.4, 1417, 9.3), (0.6,
1307, 9.4), (0.8, 1235, 9.5), (1, 1163, 9.6) \label{fig:Adiabatic-Radiative}}
\end{figure}

Heat is normally transferred out of the planet's envelope by convection
in the interior where the adiabatic gradient is lower than the radiative
one, and by radiation in the upper layers where the opacity is lower
and the converse is true. We find that this radiative-convective boundary
happens at deeper levels, larger temperatures, and lower local entropies,
as the amount of water+ices in the envelope decreases (see Fig. \ref{fig:Adiabatic-Radiative}).
The variation in opacity and pressure of this boundary is at least
an order of magnitude and decreases with increasing water+ice content
(from 4600 bars for a solar composition to 138 bars for a water/ice
envelope). In addition, with increasing age this boundary happens
at a similar local entropy which means it moves deeper (higher pressures)
as the planet cools in time. Below the boundary, the envelope is fully
adiabatic and the values for opacities are less important, as long
as they do not preclude the envelope from being convective. This means
that the extrapolation of the opacities is most important up to several
kilobars ($\sim$5000 bars) and a few thousand kelvins ($\sim$2000K)
for these warm sub-Neptune planets.

\begin{figure}
\begin{centering}
\includegraphics[width=1\textwidth]{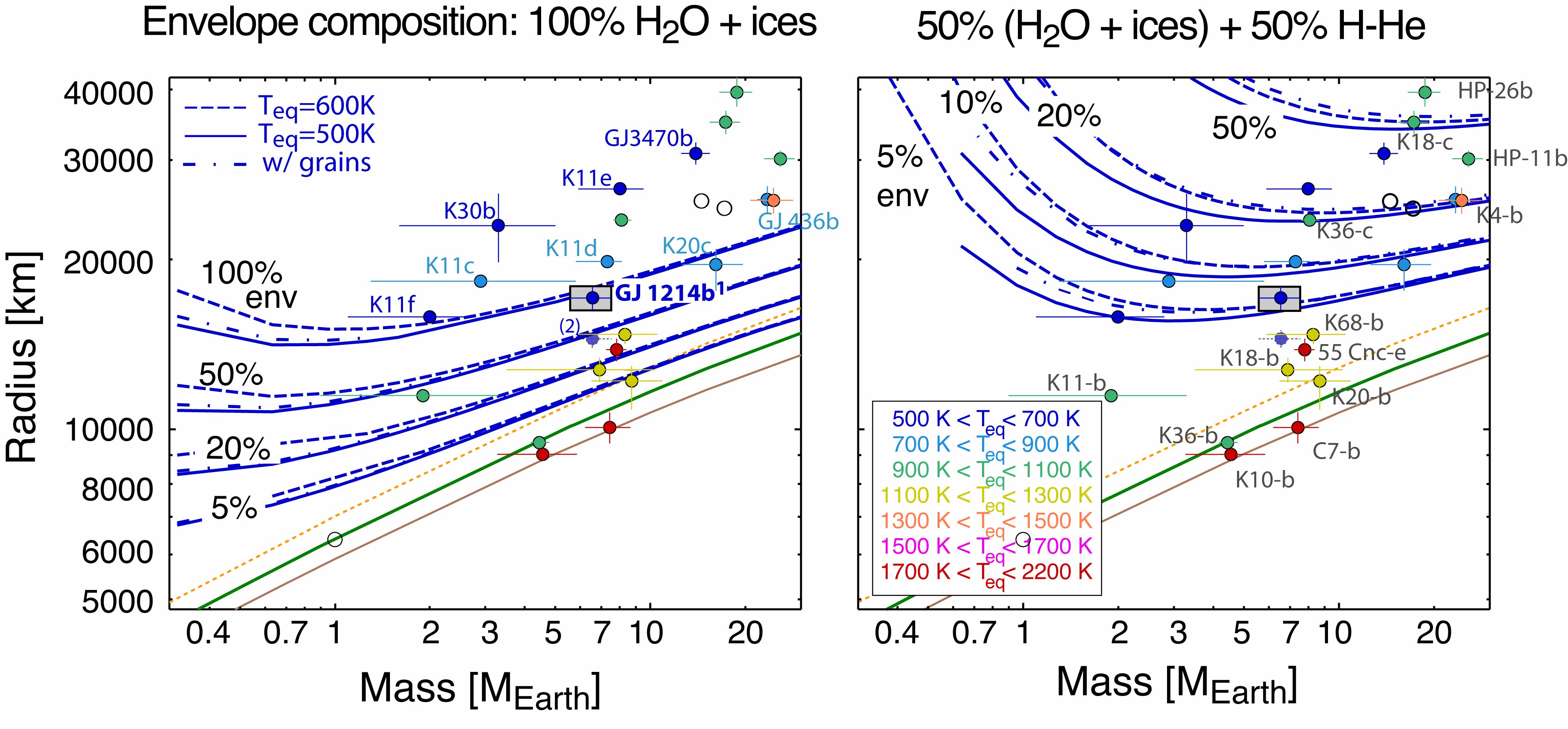}
\par\end{centering}

\caption{Mass-Radius relationships for sub-Neptunes. The relationships between
mass and radius for planets with an Earth-like nucleus below envelopes
of varying mass fraction (100, 50, 20, 10 and 5\%) are shown for a
grain-free atmosphere at $T_{eq}=500$ K (\emph{solid blue}), and
$T_{eq}=600$ K (\emph{dashed blue}), and a grainy atmosphere at $T_{eq}=500$
K (\emph{dash-dot blue}). Two envelope compositions are shown: 100\%
H$_{2}$O/ices (\emph{left}) and with 50\% (H$_{2}$O/ices)+50\% H/He
(\emph{right}). These MR relationships apply only to the planets GJ
1214b, Kepler-11e, Kepler-11f, Kepler-30b and GJ 3470b as their equilibrium
temperatures are $\sim$560K, $\sim$650 K $\sim$575 K, $\sim$ 600
K and almost $700$ K, respectively. The MR relationships shown correspond
to an age of 4.6 Gyr. Planets are color coded by their equilibrium
temperatures (calculated for an albedo of zero and an atmospheric
redistribution factor of 1/4) . Uranus and Neptune are shown for reference.
The mass-radius relationship for three rocky compositions are shown:
an Earth-like composition (\emph{green}), a Mercury-like -- enriched
in iron with respect to Earth with an iron to silicate ratio 6 times
that of Earth -- (\emph{brown}), and one voided of iron completely
(pure magnesium-silicate oxides) (\emph{orange}). The latter shows
the limit above which a planet has to have volatiles and cannot be
completely rocky. \label{fig:MR}}
\end{figure}

The effect of grain opacity is shown in Fig. \ref{fig:MR} where we
present the results for the transit radius corresponding to two different
envelope compositions: a) 100\% H$_{2}$O, b) 50\% H$_{2}$O+ 50\%
H/He, while also changing the proportion of envelope to Earth-like
nucleus. The effect of grains (dashed-dotted lines) is most noticeable
for low molecular weight atmospheres, and is negligible for water-dominated
envelopes. This is because for water dominated atmospheres the gas
opacities are already high ($\sim$10 g/cm$^{2}$) and comparable
to the grain opacities (within a factor of $\sim$0.5 dex), while
for H/He dominated atmospheres the gas opacity is 0.1-1 g/cm$^{2}$,
one order of magnitude smaller than with grains (compare solid and
dotted pink lines in Fig. \ref{fig:opacities}).

It is important to note that the mass and radius data for GJ 1214b
are consistent with a pure H$_{2}$O/ices composition (see Fig. \ref{fig:MR}-left),
regardless of the uncertainty in age, as contraction is negligible
for water dominated atmospheres. However, this composition is unlikely
to exist. The condensation temperature of water and ices is much lower
than that of rocks, so that during condensation some refractory material
should have condensed out of the solar nebula before the bulk of the
water and ices did, entailing the existence of some amount of rocky
material in this planet. This in turn, implies the presence of a material
lighter than water as well, so as to offset the high-density character
of the refractory material and fit the radius of the planet. The most
obvious component is H/He because of its abundance in astrophysical
objects, although another possibility is outgassed hydrogen \citep{Rogers_Seager:GJ1214b}. 

The effect of temperature is very small for super-Earths but modest
for sub-Earths (see Fig. \ref{fig:MR}). It is in fact, comparable
to the effect of grains, which is more important for low molecular-weight
envelopes. Relevant to GJ 1214b we can quantify the effect of equilibrium
temperature by noting that a $100$ K increase in equilibrium temperature
(that translates to an increase of $\sim$200 K at 10 bars) increases
the radius of the planet by only $\sim$2\%. The temperature effect
is small as long as the species in the envelope do not change phase
with different equilibrium temperatures. 

\begin{figure}
\begin{centering}
\includegraphics[width=1\textwidth]{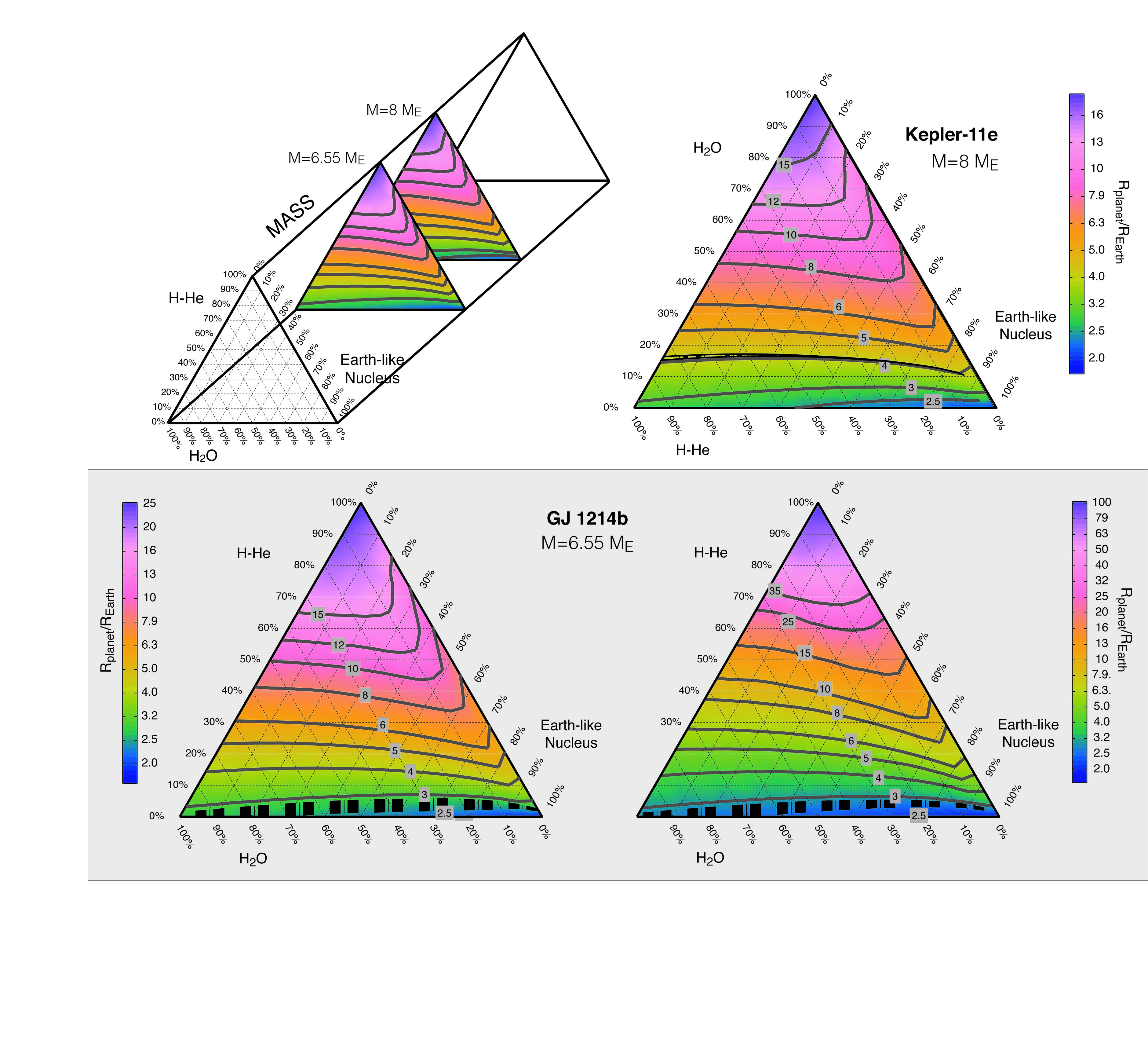}
\par\end{centering}

\caption{Ternary Diagrams for GJ 1214b and Kepler-11e. These triangular diagrams
relate the composition $\chi$ in terms of earth-like nucleus fraction,
water+ices fraction, and H/He fraction to total mass, to the radius
for a specific planetary mass. Each vertex corresponds to 100\%, and
the opposite side to 0\% of a particular component. The color bar
shows the radius in terms of Earth-radii, and the grey lines are the
isoradius curves labeled in terms of Earth-radii. The collection of
ternary diagrams for a range of planetary masses forms a triangular
prism (\emph{top left}). The black band shows the compositions constrained
by data for GJ 1214b for a grain-free envelope \emph{(bottom left)},
and a grainy envelope \emph{(bottom right)}, and Kepler-11e for a
grain free envelope \emph{(top right)} as projected onto the planetary
mass $M$ from the ternary diagrams at $M+\Delta M$ and $M-\Delta M$
(where $\Delta M$ are the uncertainty values taken from the observational
data). }

\label{fig:ternary}
\end{figure}

More systematically, we ran the internal structure model to span all
possible compositions for the envelope between the two extremes of
solar (Z$\mathrm{_{ices}}$=0.01) to 100\% H$_{2}$O/ices (Z$\mathrm{_{ices}}$=1),
and varying amounts of rocky to envelope ratios. We show the results
at a nominal age of 4.6 Gy for GJ 1214b and at 8 Gyr for Kepler-11e
in the ternary diagrams that relate Earth-like nucleus, water/ices
and H/He by mass (see Fig. \ref{fig:ternary}). Each ternary diagram
corresponds to a specific planetary mass, and every point in the ternary
diagram depicts one unique composition. These ternary diagrams are
equivalent to the (x,y,z) plane where x+y+z=1, and x,y,z > 0. The
values for the transit radius are shown in color in terms of earth
radii and the lines of constant radii are labeled. There are a few
important aspects to note from the results contained in these ternary
diagrams:
\begin{itemize}
\item The presence of H/He considerably increases the transit radius. We
find that all detected low-mass planets with a measured mass, that
have an envelope, and happen to have an equilibrium temperature warmer
than 500 K: Kepler 11b,c,d,e,f, Kepler-18b, Kepler-20b, 55Cnc-e, Kepler-68b,
Kepler-36c and Kepler-30b have a radius no larger than five times
that of the Earth. This suggests (see isoradius lines in Fig. \ref{fig:ternary})
that the H/He content is limited to less than $\sim$20\% by mass
for hot sub-Neptunes (less than 10$M_{E}$). In fact, if we remove
Kepler-11e and Kepler-30b from the list of planets, we find that the
rest of the low-mass transiting planets so far, with a measured mass,
have a maximum of 10\% by mass of H/He. Given the bias towards measuring
bigger masses, it remains to be determined if there is a population
of planets hidden in the Kepler candidates with radius 4-5 R$_{\mathrm{E}}$
that have more H/He content. 
\item The radius is most sensitive to the amount of H/He and much less to
the amount of H$_{2}$O/ices and rocky nucleus. This is seen from
how parallel the lines of constant radius are to increasing amounts
of H/He content. This means that with a radius measurement and just
some knowledge that the planetary mass ranges between 5-10 $M{}_{\oplus}$,
it is possible to estimate the H/He content of the planet. Conversely,
even with perfect data for mass and radius, it is not be possible
to estimate the amount of water/ices or refractory material, as they
trade-off quite efficiently. 
\item The effect of the presence of grains is non-linear and is most noticeable
for planets with large contents of H/He. 
\end{itemize}

\begin{figure}
\begin{centering}
\includegraphics[width=0.6\textwidth]{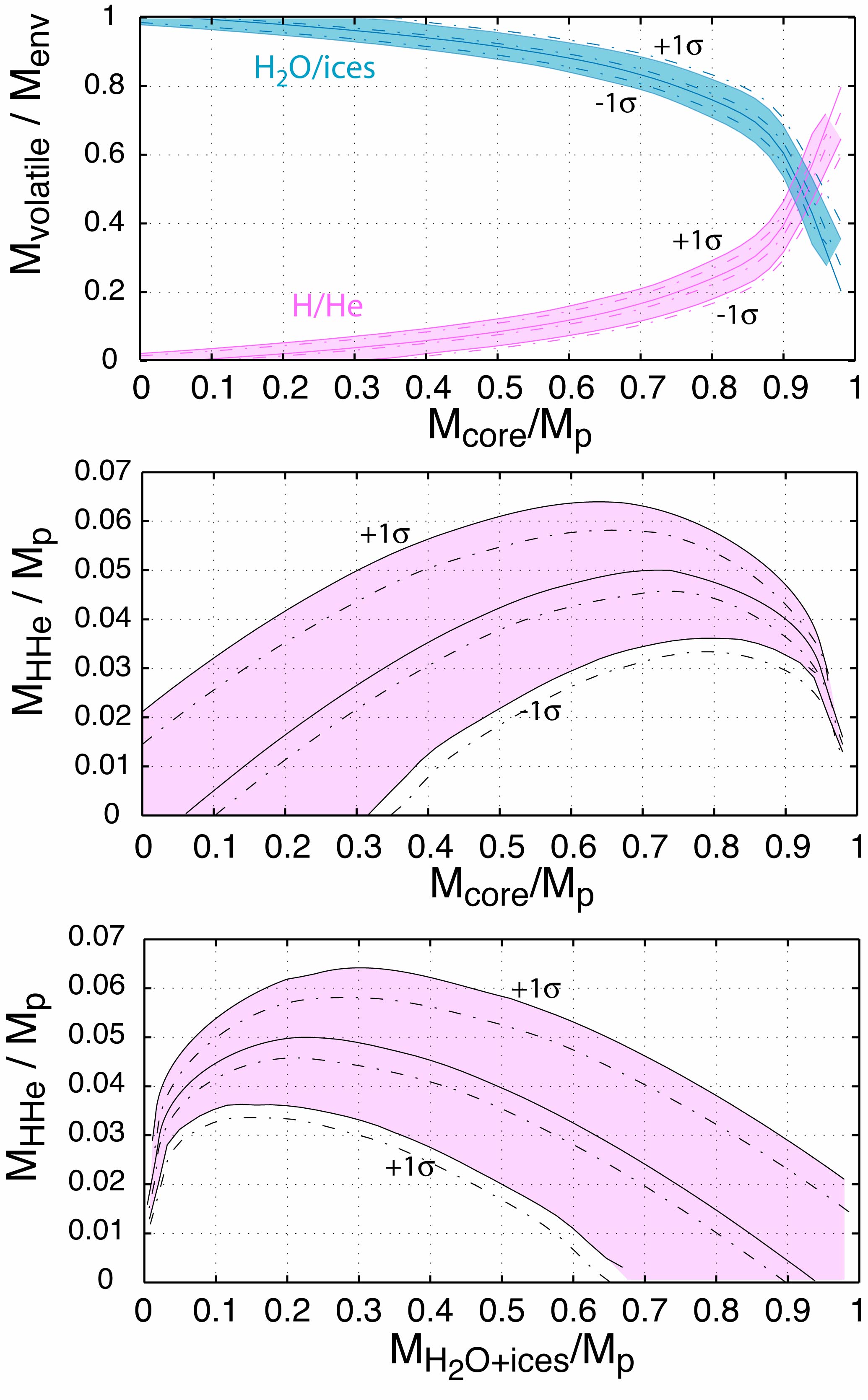}
\par\end{centering}

\caption{Composition of GJ 1214b. We fit the mass and radius of GJ1214b including
the one-sigma uncertainty to estimate the content of H/He, H$_{2}$O+ices,
and rocky nucleus. Each of the three set of lines corresponds to the
combinations $M+\Delta M,\, R-\Delta R$, $M,\, R$, and $M-\Delta M,\, R+\Delta R$,
for an old age of the system of 10 Gyr (solid lines, and shaded region)
and a younger age of 3 Gyr (dash-dotted lines). \emph{Bottom:} trade-off
between the bulk amount of H/He and water+ices by mass; \emph{middle:
}trade-off between the bulk amount of H/He and rocky nucleus by mass;
\emph{top: }proportion of H-He and water+ices in the envelope as a
function of rocky nucleus. }

\label{fig:tradeoff}
\end{figure}

The possible compositions for GJ 1214b that take into account the
uncertainty in mass and radius are shown with a black band in the
bottom panel of Fig. \ref{fig:ternary}. It is clear that this planet
has less than 10\% by mass of H/He, but that it can have a wide range
of compositions because of the trade-off between water/ices and rocky
nucleus. Another way to show the results is depicted in Fig. \ref{fig:tradeoff},
where the trade-off between bulk H/He and rocky nucleus (middle figure)
or bulk H/He and H$_{2}$O/ices (bottom figure) is shown. This can
be translated to the content of H/He and H$_{2}$O/ices in the atmosphere
(top of Fig. \ref{fig:tradeoff}). As the amount of solid core increases,
the percentage of H/He in the envelope increases while that of water
decreases. This translates to a bulk H/He content that increases as
the solid core increases up to a point where it decreases again. The
maximum amount of bulk H/He the planet may have happens in conjunction
with some water in the envelope. 

We show the range in compositions of the envelope admitted by the
data at the two limiting ages of 3 Gyr (dash-dotted lines) and 10
Gyr (solid lines) to examine the effect of age. In general, an older
planet would admit more H/He than a younger planet for a given radius.
For GJ 1214b, the fact that the age is not well constrained does not
constitute a problem when inferring the composition of its envelope,
as the effect is small. For planets older than $\sim$1 Gyr with water-dominated
envelopes, age has an effect of less than 1\% in the inference of
envelope composition. We conclude that while the total amount of H/He
in GJ 1214b can be robustly constrained to be less than 7\% by mass,
the data admits almost all possible compositions for the envelope
at any given age. In the scenario of a solar metallicity envelope
(H/He + z=0.01), we find the data constraints its content to be $\sim3$\%
by mass. 

According to their transmission spectra, \citet{Bean:GJ1214b,Bean:GJ1214b:2011}
suggest an atmosphere of more than 70\% water. If we assume that the
upper atmosphere has the same composition as the envelope below, this
range would slightly narrow the composition of the planet to have
a rocky component of less than 90\% by mass (see top panel of Fig.
\ref{fig:tradeoff}). In summary, because of the large trade-offs
between refractory material and water/ices, even with spectroscopic
measurements and the assumption that the atmosphere is well-mixed,
it is not possible to sufficiently narrow the refractory or water/ices
composition of the planet.

\paragraph{Previous studies }

Our maximum content for H/He agrees with both \citet{Rogers_Seager:GJ1214b}
and \citet{Nettelmann:GJ1214b:2011} despite having different treatments.
\citet{Rogers_Seager:GJ1214b} considered three compositions: a four
layer model with H/He above an ice layer, above an Earth-like nucleus
of silicate mantle above an iron core, or a three layer model with
vapor or outgassed H$_{2}$ above an Earth-like nucleus. In their
four-layer model they find a range of $10^{-4}-0.068$ for H/He is
admitted by the data at the one-sigma level, in their three layer
model with vapor they find a range of 47-100\%, and with outgassed
H$_{2}$ a small value of only $5\mathrm{x10^{-4}}$. While we agree
on the maximum amount of H/He and water (for obvious reasons), we
find a different value for the minimum amount of water if there is
no H/He. Our calculations show a minimum value of 65\% (see bottom
Fig. \ref{fig:tradeoff}). A possible explanation for this discrepancy
is that the opacity treatment from \citet{Rogers_Seager:GJ1214b},
which uses the Planck means from molecular line data from F08, does
not extend to very water rich atmospheres. On the other hand, \citet{Nettelmann:GJ1214b:2011}
considered a similar structure to ours with a homogeneous gas envelope.
A minor difference that should not influence the results is that they
model a homogeneous rocky interior, while we consider a layered Earth-like
one below the envelope. In their models of H/He envelope above a rocky
core they claim a range of 1.3-3.4\% of H/He. They suggest that the
upper limit of H/He can rise up to 5-6\% if the envelope contains
60-90\% water in mass. In comparison, we obtain a value of 3\% of
H/He envelope at 3 Gy, and also obtain a maximum amount of H/He by
adding 80-90\% by mass of water to the envelope (corresponding to
25\% of water by total planetary mass). We suggest that this small
difference may come from different opacity values as well. 

In addition, we find that our results are robust to reasonable variations
of thermal inertia of the planet including different radioactive heat
production or heat capacity of the Earth-like nucleus. The lower boundary
heat flux entering the envelope is $\dot{L}_{\mathrm{sol}}=\dot{\epsilon}_{\mathrm{rad}}+C_{v}dT/dt$,
where $\dot{\epsilon}_{\mathrm{rad}}$ is the radioactive heat production,
and $C_{v}$ is the heat capacity. We used a chondritic value for
the heat generation ($2\times10^{20}$ J/s/g) which is a factor of
$\sim2$ lower than Earth's bulk silicate value, and a heat capacity
of $7\times10^{7}$ J/K/g which is appropriate for the Earth \citep{Stacey:HeatCapacity:1981}.
By increasing $\dot{\epsilon}_{\mathrm{rad}}$ by a factor of 5 we
find a discrepancy at 3 Gyr of $\sim$2\%, and by increasing Cv by
a factor of 10 we find a discrepancy of $\sim$6\% for a planet with
a H/He envelope that makes 3\% and 20\% of the planet. Being that
H/He envelopes are the ones more susceptible to changes in temperature,
we conclude that the radius of a sub-Neptune planet is not very sensitive
to the thermal evolution of its rocky nucleus. This stands in contrast
to the findings by \citet{Nettelmann:GJ1214b:2011} and \citep{Lopez:K11:12}.

\subsection{Mass Loss}

It is well recognized that atmospheric escape may play an important
role in highly irradiated exoplanets \citep{Valencia:CoRoT7b}, and
GJ 1214b is no exception. Although a detailed study is beyond the
scope of this paper, we can estimate the order of magnitude effect
of atmospheric escape on GJ 1214b. Starting from the commonly-used
energy limited escape formulation \citep{Watson:AtmEsc:1981}, the
mass lost per unit time of a planet of mass $M$ is 

\begin{equation}
\dot{M}=\pi\epsilon R_{\mathrm{XUV}}^{2}R\, F_{\mathrm{XUV}}/GMK_{\mathrm{tide}},\label{eq:massloss}
\end{equation}
where $R_{\mathrm{XUV}}$ is the radius at which the bulk of the X-ray
and extreme UV (XUV) flux is absorbed, $R$ is the radius below which
molecules are bound to the planet, $F_{\mathrm{XUV}}$ is the XUV
flux at the planet's location, $G$ is the gravitational constant,
$K_{\mathrm{tide}}$ is a correction factor that takes into account
that the molecules only need to reach the Roche lobe before they escape
\citep{Erkaev:RocheLobe:07}, and $\epsilon$ is the heating efficiency
defined as the ratio of the net heating rate to the rate of stellar
energy absorption. One conservative, simple and commonly used approach
is to assume $R_{\mathrm{XUV}}\sim R$. In reality the height at which
the planet absorbs X-rays and the extreme UV are different and also
larger than the planetary radius \citep{Lammer:AtmEscXUV:03}. By
adopting the assumption, we can simplify Eq. \ref{eq:massloss} to
$\dot{M}=3\epsilon F_{\mathrm{XUV}}/4G\rho K_{\mathrm{tide}}$ where
$\rho$ is the density of the planet, which increases with time as
the planet loses mass. Mass loss progresses from fast early on, to
slow as time increases, due to two facts: (1) the lighter outer regions
get stripped away leaving a denser planet from which molecules have
a harder time escaping, and (2) the XUV flux from the star decreases
with time.

The most unconstrained parameter, and where most of the physics is
hidden in the mass loss equation is the heating efficiency, although
common values range between 0.1-0.4. Finally, it is important to know
how the XUV flux of the star has varied overtime, and while GJ 1214
is considered to be a quiet star presently \citep{GJ1214b}, being
a low-mass M star, it most likely had an active period early on. We
implement the model of XUV flux proposed by \citet{Ribas:XUV:09}.
The XUV luminosity starts in a saturated phase after which it drops
off as a power law function of age. The saturation phase duration
($t^{*}$) depends on the type of star, as seen by its bolometric
luminosity. If we focus on a conservative estimate we can further
simplify the mass loss equation by assuming that the planet loses
mass at the present density held constant. This is obviously an idealization
and a lower bound for estimating the amount of mass lost, since at
a young age planets are lighter and less capable of binding their
upper atmospheres. We also set $K_{\mathrm{tide}}=1$. The expression
for the XUV flux \citep{Ribas:XUV:10} is

\begin{equation}
 F_{\mathrm{XUV}} = \begin{cases}
 \quad 4.04\times10^{-24}\,L_\mathrm{{bol}}^{0.79} \, a^{-2} \quad  \mathrm{ (erg \, /s/cm^{-2})} & \text{ if $t_9<t_9^*$}, \text{ and} \\
\quad 29.7 \, t_9^{-1.72} a^{-2} \quad \mathrm{ (erg \, /s/cm^{-2})} &  \text{ if $t_9>t_9^*$}
\end{cases}
\end{equation}

\[
\]
where $t_{9}^{*}=1.66\times10^{20}\: L_{\mathrm{bol}}^{-0.64}$ in
Gyr. We use a value of $L_{\mathrm{bol}}=0.00328\, L_{\mathrm{Sun}}$\citep{GJ1214b},
and obtain a saturation phase duration of $2$ Gyr for GJ 1214b. We
calculate a mass loss between 100 Myr and 3 Gyr of 0.6 $M_{\mathrm{E}}$
and 2.5 $M_{\mathrm{E}}$ for a heating efficiency of 0.1 and 0.4,
respectively. This corresponds to a planet losing 9\% or 27\% of its
mass respectively. This will affect the composition and structure
of the planet. This is most important when trying to assess the origin
of the planet and the stability of an envelope. \citet{GJ1214b} estimated
through a hydrodynamic calculation that it would take 700 Myr to lose
an envelope of H/He that makes 5\% of the planet's mass. According
to our simple calculation, the current flux at the planet's semi-major
axis is 39 W/m$^{2}$, and the present mass loss rate is $ $$2.4\times10^{8}\epsilon$
kg/s or $\sim1.25\epsilon$ Earth-masses per billion years. If the
heating efficiency is close to 1, then a modest envelope ($\gtrsim0.2\, M_{p}$)
may be stable for a billion year timescale. Without a detail calculation
of atmospheric escape that includes the effects of a mixed atmosphere,
it is unclear how stable or vulnerable a thin envelope may be. On
the other hand, our simple calculation more robustly suggests that
the compositional cases where GJ 1214b has a modest envelope, seem
to be stable. Therefore, while atmospheric escape might have been
significant in the past it appears to be moderate at present for GJ
1214b.

\subsection{Comparison to Kepler-11}

A good starting point to compare low-mass planets is GJ 1214b, because
it is the coolest volatile planet and lies right at the threshold
of a pure water mass-radius relationship. This means that any of the
volatile planets over 1 Gyr old with a radius comparable to or larger
than GJ 1214b necessarily has H/He. Even though the planets shown
in Fig. \ref{fig:MR} have different ages, they are all older than
1 Gyr, with CoRoT-7b being the youngest (1.2-2.3 Gyr \citep{Leger:CoRoT-7b}),
so that the MR relationships apply. In fact, most of the transiting
planets with known mass are older than the solar system, so the inferred
amount of H-He from a younger age (of 4.6 Gyr) would be a minimum.

From Fig. \ref{fig:MR} we infer that Kepler-11f also has some H/He
in its envelope, despite its very low mass of $2.0_{-0.9}^{+0.8}M_{\oplus}$
\citep{Lissauer:K11:2013} as its radius stands above the pure-water
relationship adequate for its equilibrium temperature. In fact, because
of the behavior of the mass-radius relationships for volatile compositions
that flare out towards low masses, both planets Kepler-11f and GJ
1214b could have the same composition. This flaring effect is due
to the fact that low-mass planets have low gravities that do not bind
efficiently their volatile envelopes.

Furthermore, we focus on Kepler-11e as this planet is as cool as GJ
1214b but its radius is 1.6 times larger. We obtain all possible compositions
for Kepler-11e (top left Fig. \ref{fig:ternary}) with the new reported
data in \citet{Lissauer:K11:2013} and find that the minimum amount
of bulk H/He is 10\% and the maximum is 18\% by mass (an improvement
from the old reported radius \citep{Lissauer:Kepler-11} that yielded
10-25\% content) . Being that this planet is the largest and coolest
of the transiting super-Earths, it means that all other detected volatile
planets have less than 20\% bulk H/He. In fact, we find that all volatile
super-Earths discovered so far have less than 10\% H/He by mass, comparable
to Uranus and Neptune \citep{Hubbarb&McFarlane:1980}, except for
Kepler-11e and Kepler-30b. We find the latter to have between 5-15\%
H/He. This also brings into light that the solar system trend of decreasing
H/He with heliocentric distance for the gaseous planets \citep{Hubbarb&McFarlane:1980}
does not apply to the Kepler-11 system. 

A study by \citep{Lopez:K11:12} investigated the possible compositions
for each planet of the Kepler-11 system with an evolutionary model
and connected it to atmospheric escape histories. For the structure
part, they considered the envelope to be made of an outer layer of
H/He above an interior water layer. They use a non-grey model for
their atmosphere and opacities at 50 times solar for all their models.
For water-less worlds they report that present day (at 8 Gyr) inventories
of H/He are 17.2\% for Kepler-11e, and less than $\sim$8\% for all
others planets. This stands in excellent agreement with our results
considering that the amount of H/He would increase somewhat once they
take into account the one-sigma uncertainty in masses and radii. 

Placing constraints on the amount of H/He helps validate formation
models. According to our model, GJ 1214b must have formed rather early,
when there was still enough H/He in the solar nebula. In addition,
multi-planet systems pose an additional constrain, which is to explain
either the trend or lack thereof of H/He content with heliocentric
distance. The latter is the case of Kepler-11, with planet e having
to have at least 10\% of H/He and up to 18\% at 42 $R_{\mathrm{Sun}}$,
and neighboring planets d at 34 $R_{\mathrm{Sun}}$ and low-mass planet
f at 44 $R_{\mathrm{Sun}}$ with at most 10\% H/He. A study by \citep{Ikoma:Kepler11:2012}
investigates the formation of single low-mass planets with H/He envelopes
by invoking in-situ accretion that they then apply to Kepler-11. While
they do not consider H/He+H$_{2}$O mixtures for envelope which would
have an effect on the rate of cooling and accretion due to higher
opacities, they explain the H/He content of most of the planets in
the system. It remains to be shown how accretion of low-mass multiple
planets can acquire envelopes that also have water/ices.

\section{Summary and Conclusions }

To assess the bulk composition of low-mass, low-density exoplanets
and specifically GJ1214b, we ran a comprehensive suite of internal
structure and evolutionary models with a proposed prescription for
opacity values that span from solar to about 450 times solar -- corresponding
to a composition of 100\% H$_{2}$O/ices. 

Given that the opacities tables that are commonly used by internal
structure models are only known at discrete metallicity values that
do not cover all the possible envelope compositions that the sub-Neptune
planets may have, we focused our efforts in fitting these opacity
tables to an analytical function that describes the global behavior
of opacities in the pressure-temperature (P-T) and metallicity regime
(from water/ices) in which they are derived, as well as extrapolate
smoothly into higher P-T and water content space. The most important
regime for opacities for warm sub-Neptune planets (with an equilibrium
temperature $\sim$500 K) are up to $\sim$5 kbars and $\sim$2000
K, which covers the radiative-convective boundary in the envelope.
Opacities at larger pressures (with corresponding larger temperatures)
fall within the fully convective interior.

Interestingly, we find that the differences in opacity values of a
pure water/ices envelope and a 50 x solar envelope, which is one of
the most metal rich opacity tables available, and corresponds to 1/3
water/ices + 2/3 H/He are not too large, in the order of a few dex.
This means that while using it for envelopes with much larger water
contents is not consistent, it probably does not introduce a big source
of error in the results. 

We find that there is another type of degeneracy pertinent to sub-Neptune
planets that arises from the evolutionary history of the planet. Two
different envelope compositions of the same mass around the same rocky
nucleus may yield the same radius at a given age while differing in
the rest of their evolutionary tracks. This degeneracy is different
in character to the one that arises from the trade-offs between the
different compositional end-members.

We obtain the bulk composition of GJ 1214b and find that no more than
7\% of H/He is needed to explain the radius of this planet given its
mass. In addition, based on formation arguments we expect to have
some H/He present in the envelope. This is due to the fact that some
refractory material is expected to compose this planet (from the condensation
sequence). Our result is consistent with two previous studies focused
on GJ 1214b, which use different treatments for the opacities. 

More generally, we find that the radius of low-density planets with
a mass between 5-10 $M_{\oplus}$  is most sensitive to the amount
of H/He, and much less on the amount of water and rocks. On the upside,
this means that it is possible to place good constraints on the amount
of H/He in these planets, which can be used to further constraint
formation models. On the down side, it means that little can be said
about the amount of water or rocks in these planets because these
two compositional end-members trade-off very efficiently. 

For GJ 1214b and similar planets the implication is that the inference
of a water rich upper atmosphere from transmission spectroscopy studies
does not help constrain the bulk composition of the envelope and planet,
whereas an H/He dominated atmosphere would restrict the bulk composition
much more, only if we assume a homogeneous composition between the
upper atmosphere and deeper envelope. 

Furthermore, we find that almost all discovered low-mass planets --
Kepler 11b,c,d,f, Kepler-18b, Kepler-20b, 55Cnc-e, Kepler-68b, Kepler-36c
-- have a maximum H/He component of less than 10\% by mass. While
it could be that some of them have no H/He whatsoever, it seems that,
despite having much hotter equilibrium temperatures, the low-density
low-mass exoplanets share a similar trait to Neptune and Uranus of
having a few percent of H/He. The exceptions are Kepler-11e and Kepler-30b
with a range of 10-18\% and 5-15\% of bulk H/He respectively. Being
that there's is a bias towards detecting larger planets, the fact
that the majority of the low-mass planets have less than 10\% H/He
indicates that larger contents are probably not common. The tightly
packed Kepler-11 system seems to have a range of H/He that does not
vary monotonically with heliocentric distance, with planet d and f
having less H/He than planet e, which may point to more local/planet-specific
conditions determining the compositional outcome. In other words,
it remains to be explained why Kepler-11e acquired 10-18\% by mass
of H/He while simultaneously its inner and outer neighbors acquired
less. Multiplanet sub-Neptune systems with known H/He contents may
prove to be key in understanding planet formation.

Inferring the bulk composition of low-mass planets helps us clarify
the differences in nature between the solid super-Earths and the sub-Neptune
planets that share the same mass range, and also provide useful constraints
to formation and migration scenarios for this new class of planets.

\acknowledgements{}

This work was performed (in part) under contract with the California
Institute of Technology (Caltech) funded by NASA through the Sagan
Fellowship Program executed by the NASA Exoplanet Science Institute.
We thank Ignasi Ribas for his comments on the XUV fluxes of stars.
We thank Jonathan Fortney for his careful and insightful review that
has greatly increased the quality of the manuscript.

\bibliographystyle{apj}
\addcontentsline{toc}{section}{\refname}\bibliography{/Users/valencia/Documents/MyPapers/proj}

\end{document}